\documentclass[aps,twocolumn,showpacs,citeautoscript,reprint]{revtex4-2}

\usepackage{bm}
\usepackage{graphicx}
\usepackage{amsmath}
\usepackage{amsfonts}
\usepackage{siunitx}
\usepackage{bbold}
\usepackage{color}
\usepackage{tikz}
\usepackage{bigints}
\usepackage{booktabs,colortbl,array}
\usepackage{multirow}
\usepackage{dcolumn}
\usepackage{comment}

\begin{document}

\title[Impact of electron-electron interactions on the thermoelectric efficiency of graphene quantum point contacts]%
{Impact of electron-electron interactions on the thermoelectric efficiency of graphene quantum point contacts}

\author{Iri\'{a}n S\'{a}nchez-Ram\'{i}rez}

\affiliation{GISC, Departamento de F\'{\i}sica de Materiales, Universidad Complutense, E-28040 Madrid, Spain}

\author{Yuriko Baba}

\affiliation{GISC, Departamento de F\'{\i}sica de Materiales, Universidad Complutense, E-28040 Madrid, Spain}

\author{Leonor Chico}

\affiliation{GISC, Departamento de F\'{\i}sica de Materiales, Universidad Complutense, E-28040 Madrid, Spain}

\author{Francisco Dom\'{\i}nguez-Adame}

\affiliation{GISC, Departamento de F\'{\i}sica de Materiales, Universidad Complutense, E-28040 Madrid, Spain}


\begin{abstract}

Thermoelectric materials open a way to harness dissipated energy and make electronic devices less energy-demanding. Heat-to-electricity conversion requires materials with a strongly suppressed thermal conductivity but still high electronic conduction. This goal is largely achieved with the help of nanostructured materials, even if the bulk counterpart is not highly efficient. In this work, we investigate how thermoelectric efficiency is enhanced by many-body effects in graphene nanoribbons at low temperature. To this end, starting from the Kane-Mele-Hubbard model within a mean-field approximation, we carry out an extensive numerical study of the impact of electron-electron interactions on the thermoelectric efficiency of graphene nanoribbons with armchair or zigzag edges. We consider two different regimes, namely trivial and topological insulator. We find that electron-electron interactions are crucial for the appearance of interference phenomena that give rise to an enhancement of the thermoelectric efficiency of the nanoribbons. Lastly, we also propose an experimental setup that would help to test the validity of our conclusions.

\end{abstract}

\maketitle

\section{Introduction}

Understanding the mechanisms behind heat-to-electricity conversion is crucial for developing efficient and functional thermoelectric devices. Wistfully, the lack of efficiency has been a burden since the dawn of thermoelectricity research, hindering the development of devices with widespread applications. During the last three decades, theoretical~\cite{Hicks1993} and experimental~\cite{Boukai2008} evidences for the enhancement of thermoelectric properties in the nanoscale have arisen. Thereupon, nanotechnology postulates as a promising framework for research and development of the aforementioned devices.

One way to quantify the thermoelectric efficiency of a certain material or device is the dimensionless figure of merit, defined as  $ZT=S^2\sigma T/\kappa$, where $S$ is the Seebeck coefficient and $\sigma$ and $\kappa$ are the electric and thermal conductivity, respectively~\cite{Goldsmid2010}. For example, being $ZT=4$ considered as a suitable value for widespread applications, efficient bulk materials display values for $ZT$ of the order of unity~\cite{Biswas2012}, while figures of merit around $ZT=2.4$ have been obtained for thin-film superlattices even at room temperature~\cite{Venkatasubramanian2001}. There are a plethora of mechanisms by which $ZT$ can be increased in the nanoscale, from strong phonon scattering~\cite{Khitun2000} to nanopore tailoring in graphene nanoribbons~\cite{Chang2012}, to name a few. One of these mechanisms
%, and the one in which we are interested in this work, 
consists in the enhancement of the thermoelectric response through interference-related phenomena. This mechanism has been widely investigated in molecule heterojunctions~\cite{Bergfield2009}, quantum dots~\cite{Karlstrm2011} and other nanostructures \cite{Cortes2016, Chico2017}. Besides, interference between resonant and non-resonant processes leads to the appearance of Fano resonances~\cite{Fano1961} in the transmission coefficient, which are predicted to have a positive impact on the efficiency~\cite{GarcaSurez2013}. In this work, we explore how such interference phenomena can be also related to electron-electron interactions. 

Therefore, in order to uncover the impact of electron-electron ($e-e$) interactions in the thermoelectric efficiency of a certain material or nanostructure, we shall first understand its role in the occurrence of interference phenomena in electron transmission through these systems. The effect of $e-e$ interactions on electronic transport in two-dimensional topological insulators (2DTIs) and trivial insulators, which can be used in quantum point contacts, is still poorly understood. Despite that, it is a topic of interest in current research (e.g. see Ref.~\cite{Novelli2019}), where the breakdown of quantized conductance in 2DTIs is addressed by employing a mean-field approach to the Kane-Mele model~\cite{Kane2005}, successfully reproducing experimental results for quantum wells~\cite{Konig2007} and atomically thin crystals~\cite{Wu2018}. 

Within this framework, we present a mean-field approach which captures the interplay between $e-e$ interactions, topologically protected helical edge states, transport and thermoelectric efficiency. Moreover, we propose and consider graphene nanoribbons (GNRs) as the system of study and a possible experimental realization of this model. This choice is grounded in several arguments. To begin, because of its background knowledge and accessibility: graphene has been widely studied and understood both theoretical~\cite{Katsnelson2007} and experimentally since its discovery in 2004~\cite{Novoselov2004}. To follow, because of its transport properties: graphene stands out as a very useful material for constructing electronic devices due to its robust coherent electronic transport against disorder and modifications on its geometry~\cite{MuozRojas2006}. And last, due to the feasible experimental implementation of this work: owing to new fabrication techniques~\cite{Cleric2019}, high-quality graphene nanostructures and characterization is fairly accessible. Besides, when exposed to an in-plane magnetic field~\cite{Young2013}, GNRs are shown to display helical edge transport intrinsic to a quantum spin Hall phase (QSH), hence encapsulating the behavior of 2DTIs. Furthermore, recent studies~\cite{Kim2020} show that $e-e$ interactions can be modulated in graphene by using proximity screening. 

\indent This paper is organized as follows. In Sec.~\ref{sec:Model} we present the theoretical framework and the methods we apply for obtaining the transport and hence thermoelectric properties of  GNRs. In Sec.~\ref{sec:Results} we present our results in two steps. First, in Secs.~\ref{sec:QSH} and~\ref{sec:TBH}  we investigate the effect of $e-e$ interactions in GNRs with and without helical edge states. Second, in Sec.~\ref{sec:termo} we relate this effect to the thermoelectric efficiency of each GNR considered. Lastly, in Sec.~\ref{sec:concl} we make a brief summary of these results as a conclusion and set out an experimental proposal.
\section{System and model Hamiltonian} \label{sec:Model}
\begin{figure}[ht]
    \centering
    \includegraphics[width=0.9\columnwidth]{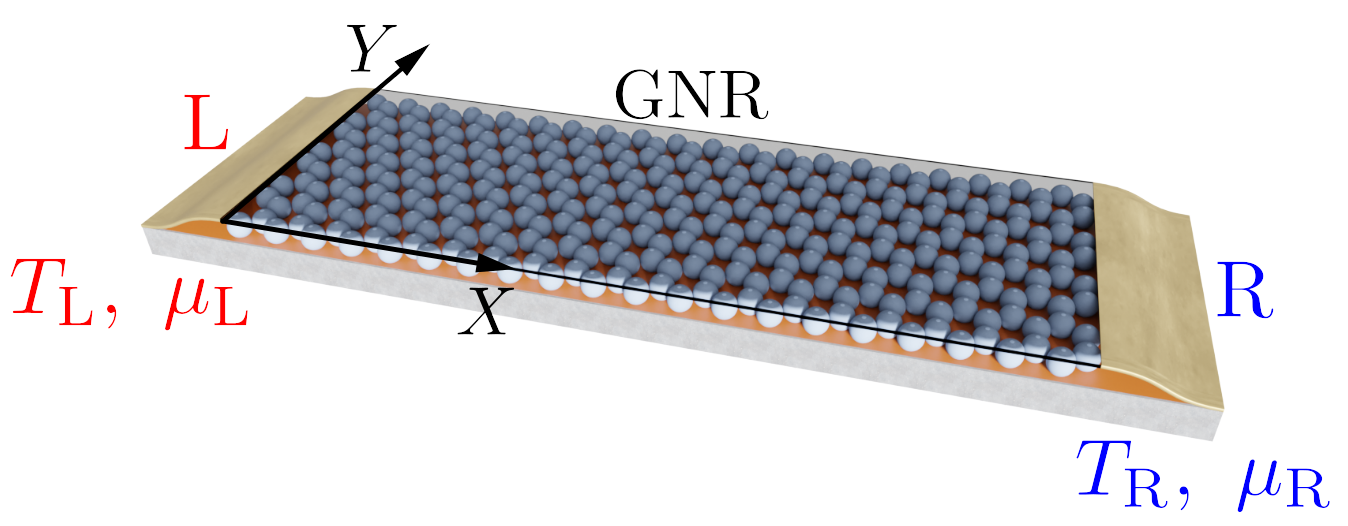}
    \caption{Sketch of the device. The GNR is connected to two ideal leads, labeled $L$ and $R$. The device is driven out of equilibrium by a temperature gradient ($T_L \neq T_R$) and/or a source-drain voltage ($\mu_L \neq \mu_R$).}
    \label{fig:setup}
\end{figure}

In this section, we present the systems under consideration (see Fig. \ref{fig:setup}) and the procedure for evaluating their thermoelectric efficiency through transport. Since the presence of point defects such as monovacancies can affect electronic properties of GNRs due to electron-defect scattering~\cite{Cui2011} and, in turn, enhance their thermoelectric response via gap opening~\cite{Kolesnikov2017}, we consider both GNRs with and without vacancies to clarify the role of $e-e$ interactions in combination with electron-vacancy scattering. 

Throughout this work we evaluate the impact of $e-e$ interactions in the thermoelectric efficiency of GNRs with different edge terminations in two scenarios, namely, when helical edge states contribute to transport and when they do not. To this end, we study the transport properties of two types of GNRs, one with zigzag edges (zGNR) and another with armchair edges (aGNR).

In order to study the electronic structure of the GNRs we consider a Kane-Mele-Hubbard model~\cite{Kane2005}. The choice of the model is justified for two reasons. On the one hand, it allows us to study the effect of local Coulomb interaction between electrons and, on the other, it enables us to easily control the appearance of helical edge modes. The Hamiltonian of this model reads~\cite{Novelli2019}
\begin{align} \label{eq:ham1}
\begin{split}
 \mathcal{H} = \; &t\sum_{\langle ij\rangle} \sum_{\alpha}c_{i\alpha}^{\dagger}c_{j\alpha}^{ } 
 +i\lambda\sum_{\langle\langle ij\rangle\rangle}\sum_{\alpha\beta}\nu_{ij}\sigma^z_{\alpha\beta}c_{i\alpha}^{\dagger}c_{j\beta}^{ }\\
 +&U\sum_{i}n_{i\uparrow}^{}n_{i\downarrow}^{}\ ,
\end{split}
\end{align}
where $c_{i\alpha}^{}(c_{i\alpha}^{\dagger})$ annihilates (creates) an electron with spin $\alpha=\uparrow,\downarrow$ at the $i$-th site of the lattice and $n_{i\alpha}^{}=c_{i\alpha}^{\dagger}c_{i\alpha}^{}$ is the number operator. The Hamiltonian (\ref{eq:ham1}) splits into three terms. The first one is the conventional nearest-neighbor tight-binding term and the sum $\langle ij \rangle$ runs over nearest-neighbors. The contribution of this term is modulated by the tunnel energy $t$, which is set to be $t=2.7\,\mathrm{eV}$~\cite{CastroNeto2009}. The second term accounts for the hoppings of electrons between second-nearest-neighbors, where the sum is intended between second neighbors $\langle\langle ij \rangle\rangle$, allowing spin-flip events due to the third Pauli $2\times 2$ matrix $\sigma^z$, acting on spin space. This term was introduced by Kane-Mele and is the responsible for the appearance of the aforementioned helical edge modes by virtue of $\nu_{ij}$. It acts as an intrinsic spin-orbit coupling introducing a factor $-1$~($+1$) if the hopping of the electron is (counter-) clockwise. The contribution of this term is modulated by the parameter $\lambda$, which we will express in units of $t$ hereafter. In the case $\lambda=0$, helical edge states disappear and we recover a standard tight-binding model with $e-e$ interactions. The last term accounts for the local repulsive Coulomb force between electrons of opposite spin and is modulated by $U$. To solve this many-body Hamiltonian we use an unrestricted Hartree-Fock approximation that we derive in the Appendix \ref{sec:MFA}.

Throughout this work, we consider only the electronic contribution to heat transport. We make this consideration on the grounds that for graphene nanostructures, phonon contribution to heat transport can be studied separated from electronic contribution~\cite{Miloevi2010} and, furthermore, at the considered temperature ($T=4$ K) it is rather negligible in GNRs~\cite{Mazzamuto2011,SaizBretn2019} and thus $\kappa_{\text{tot}}=\kappa_{\text{el}}+\kappa_{\text{ph}}\approx \kappa_{\text{el}}$. In this section we derive the figure of merit $ZT_{\text{el}} \approx ZT$, which will be used as an indicator of the thermoelectric efficiency of our systems~\cite{Goldsmid2010}. The figure of merit at low temperature then reads
\begin{align}
   ZT = \frac{S^2 \sigma T}{\kappa_\mathrm{el}}\ , 
   \label{ZT}
\end{align}
where $S$ is the Seebeck coefficient while $\sigma$ and $\kappa_e$ are the electric and thermal conductivities at a given temperature $T$. 

In the linear regime ($T_L\simeq T_R \equiv T$ and $\mu_L\simeq \mu_R \equiv \mu$), the magnitudes of interest can be expressed as ~\cite{Liu2010,GmezSilva2012,Nunez2020}
\begin{subequations}
\begin{align}
    S & = -\,\frac{1}{eT}\frac{K_1}{K_0}\ , \nonumber\\ 
    \sigma & = e^2K_0\ , \nonumber \\
    \kappa_\mathrm{el} & = \frac{1}{T}\left(K_2-\frac{K_1^2}{K_0}\right) \ ,
\end{align}
where we defined
\begin{equation} 
    K_n = \frac{1}{h}\int \mathrm{d}E\; \tau(E)\left(-\frac{\partial f}{\partial E} \right)(E-\mu)^n\ .
    \label{eq:kn}
\end{equation}
\end{subequations}
Here $\tau(E)$ is the transmission coefficient and $f(E,T,\mu)$ is the Fermi-Dirac distribution. Therefore, the figure of merit~\eqref{ZT} can be cast in the form
\begin{align}
    ZT=\frac{K_1^2}{K_0K_2-K_1^2}\ .
    \label{eq:zt}
\end{align}
It is worth stressing that this expression is valid in the linear response regime, assuming coherent electron transport.

\section{Results} \label{sec:Results}

In this section, we present the results obtained by applying the methods introduced in Sec.~\ref{sec:Model}. First, in Secs.~\ref{sec:QSH} and~\ref{sec:TBH}, we present and compare the transport properties at low energies for aGNRs and zGNRs in two regimes, namely QSH, where $\lambda, U\neq 0$ and thus topologically protected helical edge modes are present, and tight-binding Hubbard (TBH), where $\lambda=0$ and $U\neq 0$. The results were obtained close to the Fermi energy ($E\sim E_F$) for two main reasons. On the one hand, the impact of protected helical edge modes in electronic transport is greater for energies lying in the energy gap for bulk modes, and in the other, electron-vacancy interference events that enhance thermoelectric efficiency arise mainly at low energies~\cite{Orlof2013,Bahamon2010}. Afterwards, in Sec.~\ref{sec:termo} we make use of the results obtained in Secs.~\ref{sec:QSH} and~\ref{sec:TBH} to understand the interplay between Coulomb interaction, helical edge states and vacancy-induced interference phenomena in the thermoelectric response of GNRs.

\subsection{QSH regime}\label{sec:QSH}

\begin{figure}
    \vspace*{-1cm}
    \centering
    \includegraphics[width=0.8\columnwidth]{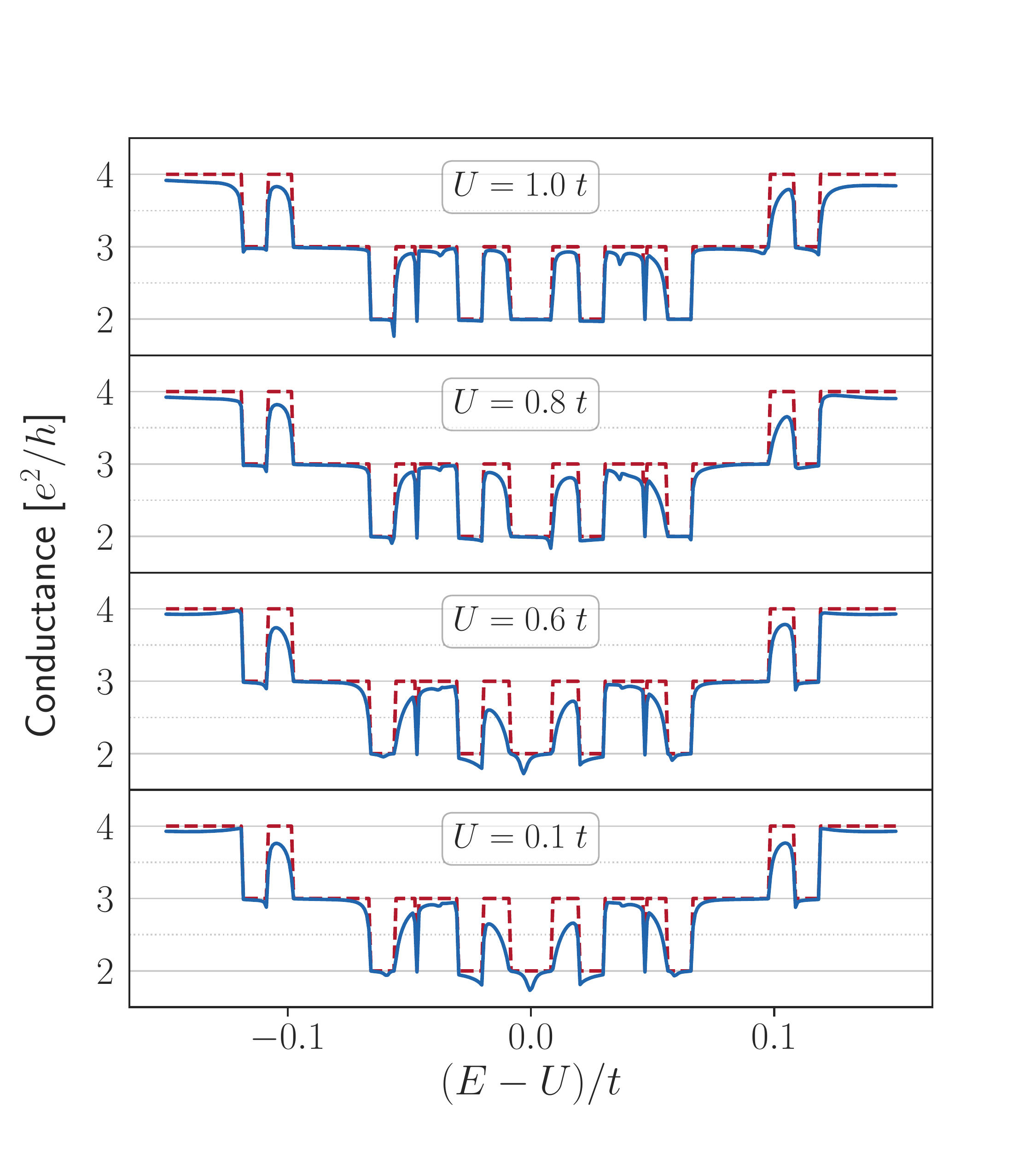}
    \vspace*{-0.4cm}
    \caption{Conductance expressed in units of $e^2/h$ as a function of $(E-U)/t$ for an aGNR of width $ W \sim 11\,\mathrm{nm}$ and length $L\sim 15\,\mathrm{nm}$ for several values of $U$ expressed in units of $t$ in the QSH regime. Dashed lines represent the conductance of pristine GNR and continuous lines show the conductance for GNR with a vacancy located at $(x,y)\approx (L/2,1.5a)$ where $a=0.25 \,  \mathrm{nm}$ is graphene lattice constant.}
    \label{fig:accon}
    \vspace*{-0.3cm}
\end{figure}

\begin{figure}[ht]
    \vspace*{-1.1cm}
    \centering
    \includegraphics[width=0.8\columnwidth]{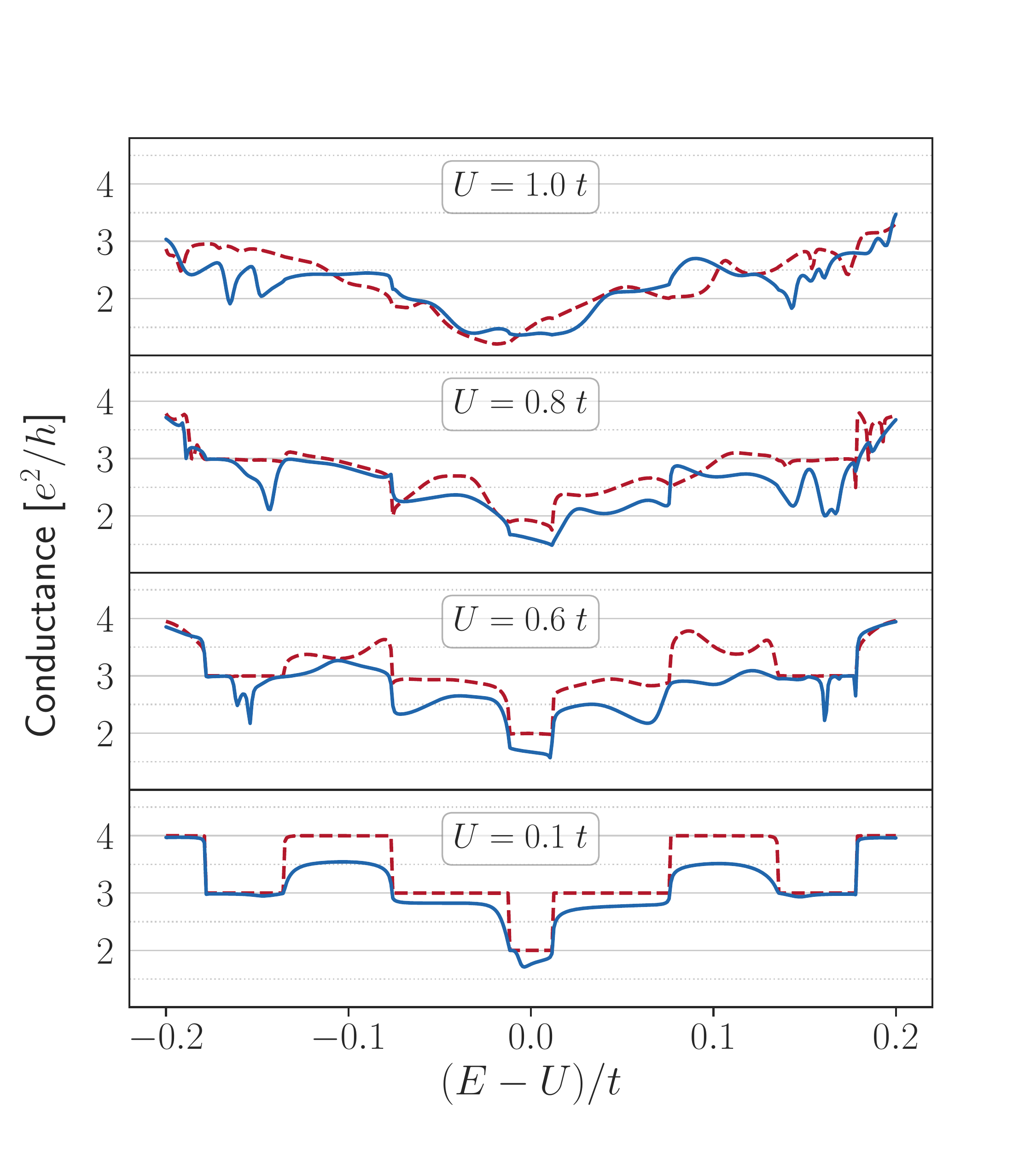}
    \vspace*{-0.6cm}
    \caption{Conductance expressed in terms of $e^2/h$ as a function of $(E-U)/t$ for a zGNR of width $ W \sim 12\, \mathrm{nm}$ and length $L\sim 15\, \mathrm{nm}$ for several values of $U$ expressed in terms of $t$ in QSH regime. Dashed lines represent the conductance of pristine ribbons and continuous lines show the conductance for ribbons with a vacancy at $(x,y)\approx (L/2,W-1.5a)$.}
    \label{fig:zzcon}
\end{figure}

In Figs.~\ref{fig:accon} and~\ref{fig:zzcon} we present the conductance as a function of energy for aGNRs and zGNRs in QSH regime. For both of them, we set $\lambda =0.09 t$ and plotted the conductance for various values of $U/t$ in the range $0.1$ to $1.0$. In those figures, dashed lines represent the conductance for pristine GNRs and solid lines stand for the conductance of GNRs with a vacancy at a distance $1.5a$ from the edge, where $a$ is the lattice constant. Vacancies were placed near the edge in order to study the interplay between topologically protected edge modes and the possible quasi-bound state expected around the vacancy~\cite{Deng2014}.

The results shown in Fig.~\ref{fig:accon} correspond to gapped aGNRs of width $W\sim 11 \, \mathrm{nm}$ ($N=51$~\cite{Wakabayashi2010}) and length $L\sim 15 \, \mathrm{nm}$. All these results present oscillations of the transmission coefficient between $\tau = 2$ and $\tau=3$ at energies in the range $U-0.1t <E < U+ 0.1t$ both for pristine and defective GNRs. We relate this oscillations with a possible hybridization between the helical edge modes and bulk modes lying near $E_F$ due to the sizable width of the GNR. This hybridization arises because the energy gap for bulk states is inversely proportional to aGNRs with $N=3p$ (with $p$ an integer)~\cite{Kimouche2015}, allowing those modes to lay near to $E_F$. Despite the aforementioned hybridization, at energies very close to Fermi energy, $E\approx U$, the transmission coefficient is $\tau(U) =  2$ for pristine GNRs owing to the two propagating edge modes arising from the Kane-Mele term as expected. 

From the curves in Fig.~\ref{fig:accon} we can make two statements. Firstly, $e-e$ interactions have a negligible impact on transmission in pristine GNRs, at least at low energies. Secondly, conductance around Fermi energy is more affected by the presence of a vacancy for small values of $U$ in the range $0.1 \leq U/t \leq 0.6$ rather than for greater values $0.8 \leq U/t \leq 1.0.$ For defective GNRs, conductance shows a resonant dip whose resonant energy varies with $e-e$ interaction strength, in accordance with~\cite{Novelli2019}. For small values of $U$, $0.1 \leq U/t \leq 0.6$, the dip is located near $E=U$ whilst for larger values it is shifted to higher energies (in absolute value). Independently of the strength of $e-e$ interactions, the transmission coefficient never reaches values under unity, namely $ \tau(E) > 1$ over the whole range of energies since the vacancy only affects the helical mode propagating at the edge where it is placed,  allowing the other one to propagate without scattering. 

\begin{figure}
    %\vspace*{-1cm}
    \centering
    \includegraphics[width=0.85\columnwidth]{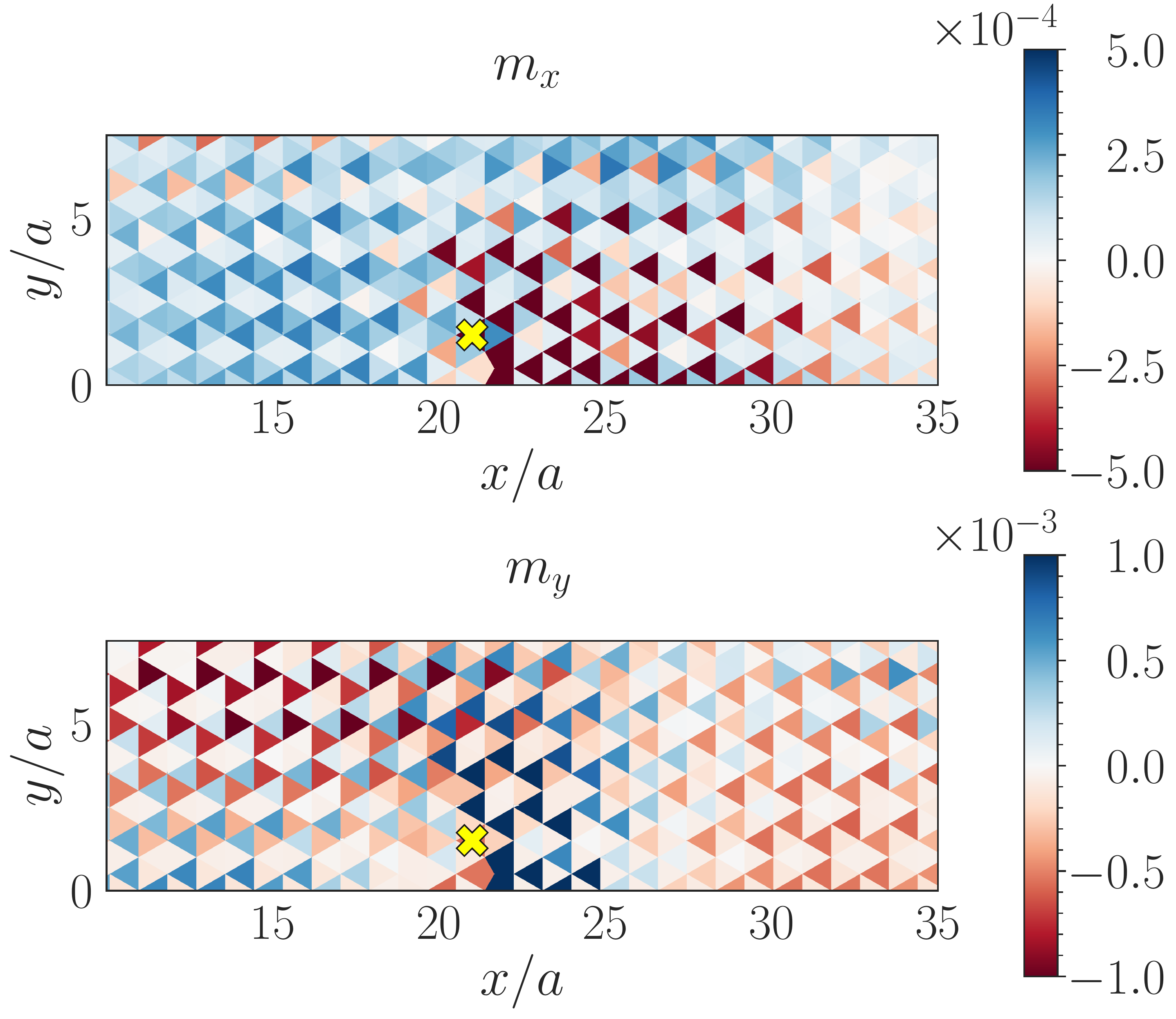}
    \caption{In-plane dimensionless magnetic moment profile at and around the vacancy obtained from Eq.~\ref{eq:eig} for an aGNR in QSH regime with $\lambda/t = 0.09$ and $U/t = 0.6$ and a vacancy located at $(x,y)=a(21,1.5)$, which is indicated with a yellow $\bm{\times}$ symbol. Triangles represent each lattice site and here we show a part of the ribbon which includes the vacancy. $m_z$ is not plotted since it was negligible ($m_z\sim 10^{-6}$).}
    \label{fig:mxac}
\end{figure}

%{\sc Aqui me he quedado}

In order to understand these results, we plot the in-plane component of the magnetic moment at the resonant energy in Fig.~\ref{fig:mxac} for an aGNR with $\lambda/t=0.09$ and $U/t=0.6$. Figure~\ref{fig:mxac} shows a clear in-plane magnetic moment localized around the vacancy and two magnetic tails spreading along the  transport $x$-direction. The spin orientation of these tails is opposite when approaching the vacancy from smaller or greater values of the $x$ coordinate. In other words, the spin orientation at the left and right sides of the vacancy is positive and negative, respectively. We can thus relate the breakdown of conductance quantization to the appearance of a nontrivial in-plane magnetic moment around the vacancy which induces backscattering through spin-flip events, as evidenced by the aforementioned magnetic tails. The mechanism behind the appearance of the resonant dips is related to the electron behavior when a vacancy is placed near the edge in QSH regime. In this regime, vacancies located at the edge act as a magnetic impurity with an in-plane magnetic momentum and an onsite energy which varies with $U$. This magnetic ordering breaks time-reversal symmetry~\cite{Lado2015,Rachel2010} an thus topological protection of the helical edge states, causing a Breit-Wigner-like resonance. The previously mentioned hybridization is also observed in Fig.~\ref{fig:mxac}, in which the magnetic moment tails arising from backscattering decay not only along the transport direction but also into the bulk, where magnetization is not trivial either, supporting our hypothesis. To sum up with, $e-e$ interactions favors the formation of in-plane magnetic nontrivial momenta around vacancies which act as magnetic impurities whose onsite energy depends on the strength of the Coulomb interaction. This pseudomagnetic impurities break down conductance quantization by inducing spin-flip events and thus backscattering in the helical edge states around the aforementioned onsite/resonant energy. The relation between the strength of $e-e$ interactions and the magnitude of destruction of then conductance quantization is not fully clear. Nevertheless, we could relate it to the shifting in energy of the pseudomagnetic impurity onsite energy, causing this energy to lie far form the gap of bulk modes, which seem to be less susceptible to these impurities, at least at those energies.

\begin{figure}[ht]
    \vspace*{-0.60cm}
    \centering
    \includegraphics[width=0.8\columnwidth]{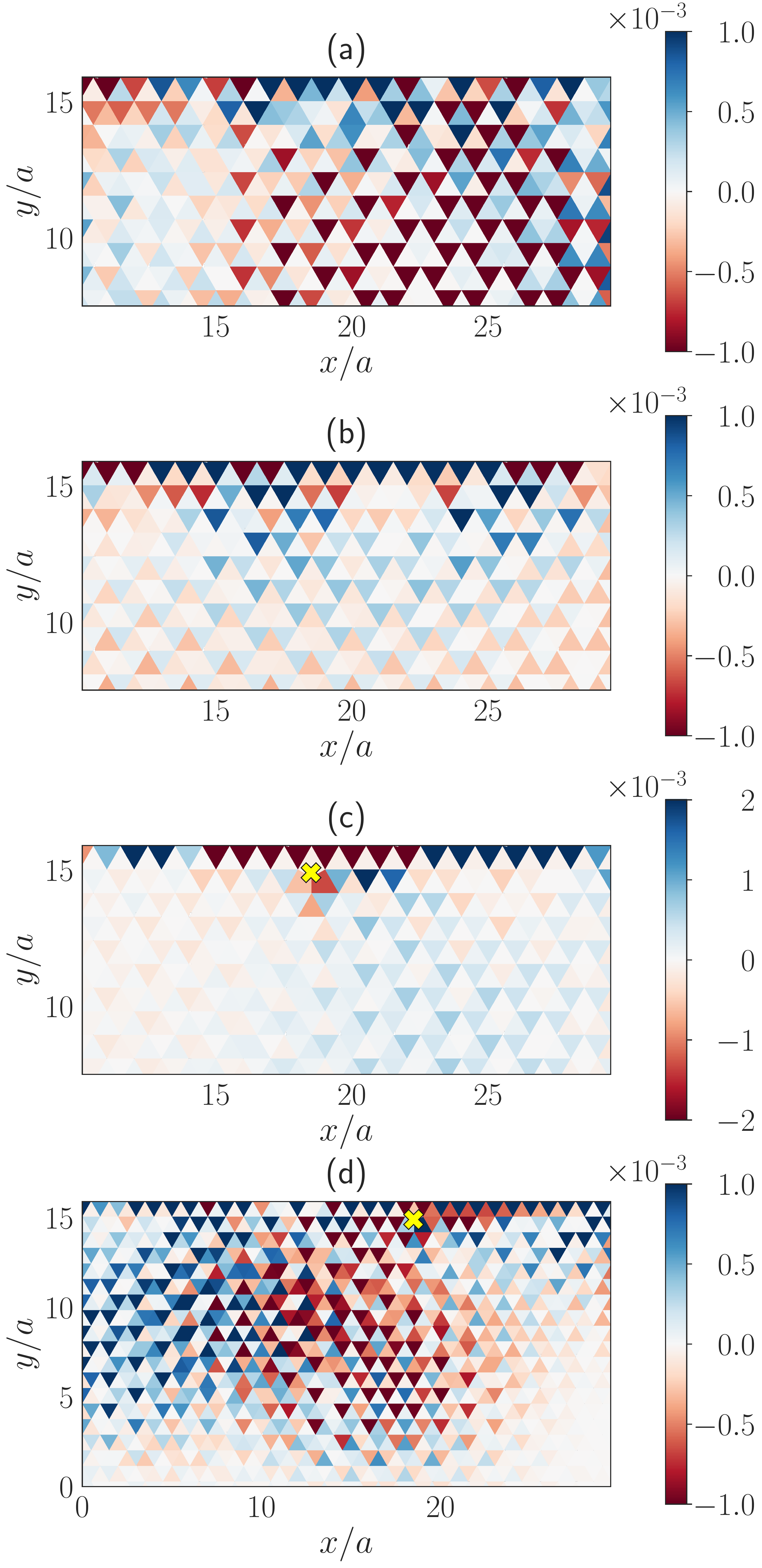}
    %\vspace*{0.10cm}
    \caption{Dimensionless $x$-component of the magnetic moment at and around the vacancy for an zGNR in QSH regime with $\lambda/t = 0.09$. (a)~Pristine zGNR with $U/t = 1.0$, (b)~pristine zGNR with $U/t = 0.6$, (c)~defective zGNR with $U/t = 0.1$ and a vacancy located in $(x,y)=a(18,15)$, and (d)~defective zGNR with $U/t = 0.6$ and a vacancy located in $(x,y)=a(18,15)$, indicated with a yellow $\bm{\times}$ symbol. Triangles in the figure represent each lattice site and here we show the region of the ribbon which includes the vacancy. $m_z$ and $m_y$ are not plotted since $m_z$ was negligible ($m_z\sim 10^{-6}$ and $m_y\sim m_x$).}
    \label{fig:mxzz}
\end{figure}

The results shown in Fig.~\ref{fig:zzcon} correspond to zGNRs of width $W\sim 12\, \mathrm{nm}$ ($N=35$) and length $L\sim 15 \, \mathrm{nm}$. In tight-binding (TB) regime, i.e., $U=\lambda=0$, zGNRs are all metallic and show localized (nondispersive) edge states which are responsible for electronic transport at low energies through bulk~\cite{Wakabayashi2010}. These edge-localized states are different in nature from the dispersive ones arising from Kane-Mele term by two means: the former ones are entirely localized within one sublattice (namely, the one forming the edge) and nondispersive, while the latter ones are dispersive and nonvanishing on both sublattices. With this in mind, we thus study here the interplay between these geometry-related localized edge states and the helical ones. The most remarkable feature of the results shown in Fig.~\ref{fig:zzcon} in comparison with those of Fig.~\ref{fig:accon} is that conductance quantization is spoiled at low energies even for pristine zGNRs. The deterioration of conductance quantization, both in pristine and defective zGNRs, is aggravated by the increase of the $e-e$ interaction strength, even completely disappearing when $U\sim t$. The main difference between the results for the conductance of defective zGNRs in comparison with those of pristine ones is the appearance of one or three asymmetric resonant dips which depend on $U$. For small values of $U$, $U/t\sim 0.1$, a resonant dip appears near the Fermi energy, where for the pristine zGNR the transmission is $\tau = 2$. This resonant dip is similar to the one arising for $U/t=0.1$ in defective aGNRs in the sense that, in both of them, the transmission never reaches values under $1$, suggesting that the responsible modes for electronic transport at low energies are the same in both types of GNR. Despite that, the zGNR resonant dip is more asymmetric, which suggests an interplay between dispersive and nondispersive edge modes that we will explore by studying the spatial profile of the magnetization at $E\sim U$ in Fig.~\ref{fig:mxzz}. This low-energy resonant dip also appears for greater values of $U$, where its strength is amplified, as evidenced by the increase in width of the resonant dip~\cite{Fano1961}. For defective zGNRs with greater $U$, another resonant dips appear at larger energies, around $E\approx U+0.15 t$. These dips are highly asymmetrical and enhanced by the $e-e$ interaction strength. These modes arise when the transmission coefficient reaches $\tau=3$, to wit, in a region where the first bulk mode is starting to propagate and hybridize with the edge modes. At the resonant energy of these resonant dips, the transmission coefficient goes from $\tau=3$ to $\tau=2$ showing the destructive impact of the vacancy on one of the propagating modes in the scattering region. 

In order to understand these results and compare them with the ones obtained for zGNRs, in Fig.~\ref{fig:mxzz} we plot the spatial profile for the magnetic moment at and around the vacancy at different energies. As in aGNRs, the magnetic moment lie mainly in-plane as expected in the QSH regime. Near the Fermi energy, for pristine samples and low $U$, there is a nontrivial density of states at the edges of the aGNR whose magnetic momentum is not uniform along the edge as shown in Fig.~\ref{fig:mxzz}(b). As expected for aGNRs in TB regime, those states are localized at the edge and decay exponentially into the bulk. For greater $U$ ($U/t=0.8,1.0$), these localized edge states are distorted and spontaneous magnetization arise for $U/t=1.0$, as shown in Fig.~\ref{fig:mxzz}(a).  We have to be cautious about the results for spontaneous magnetization in $U/t=1.0$ since it could be a spurious result of the mean-field approximation due to the fact that for larger $U$, a phase transition from QSH regime to spin density wave is expected~\cite{Rachel2010,Sorella_1992}. Our mean-field approach fails to describe this situation and thus we cannot make further hypothesis about the results for $U \geq 1$. For defective samples, there is a peak in the local density of states (LDOS) around the vacancy at the Fermi energy which displays magnetic polarization, as shown in Fig.~\ref{fig:mxzz}(c). Despite this magnetization, there are no evident spin-flip events. For larger energies, i.e., $E\approx U+0.15t$, the results are more similar to the ones obtained for aGNRs, there are evident signals of spin-flip events around the vacancy and a nontrivial magnetic moment. The main difference with aGNRs it that in the case of zGNRs, spin-flip events occur slightly inside the bulk, exhibiting the importance of hybridization in the case of zGNRs, as shown in Fig.~\ref{fig:mxzz}(d). We believe these events are the mechanism behind the appearance of resonances at these energies. Nevertheless, although the underlying mechanism is the same, there are some important differences between the two types of GNRs. In zGNRs, the modes suffering backscattering are fully hybridized and thus the lineshape of the resonant dip is asymmetric. We identify these as Fano resonances: the backscattering event undergone by the dispersive edge modes is a resonant process which takes place at the same energy as the bulk mode transmission. The vacancy is acting as a magnetic impurity just as in aGNRs but its onsite and thus resonant energy is displaced to higher values due to the nondispersive edge modes. This displacement is similar to the one happening in TB regime, where the resonant energy for vacancies located near the edge are displaced in energy from the ones located in the bulk~\cite{Xiong2010} due to the increase in the energy of the quasi-bound state localized at the vacancy. Once the resonant dips arising for defective GNRs are addressed, we link the deterioration of conductance quantization in pristine samples with the appearance of spontaneous nontrivial magnetic moments induced by the effect of Coulomb interactions in nondispersive edge states. Further analysis of this effect will be carried out through the following Subsection.

\begin{figure}[h!]
    \vspace*{-1cm}
    \centering
    \includegraphics[width=0.8\columnwidth]{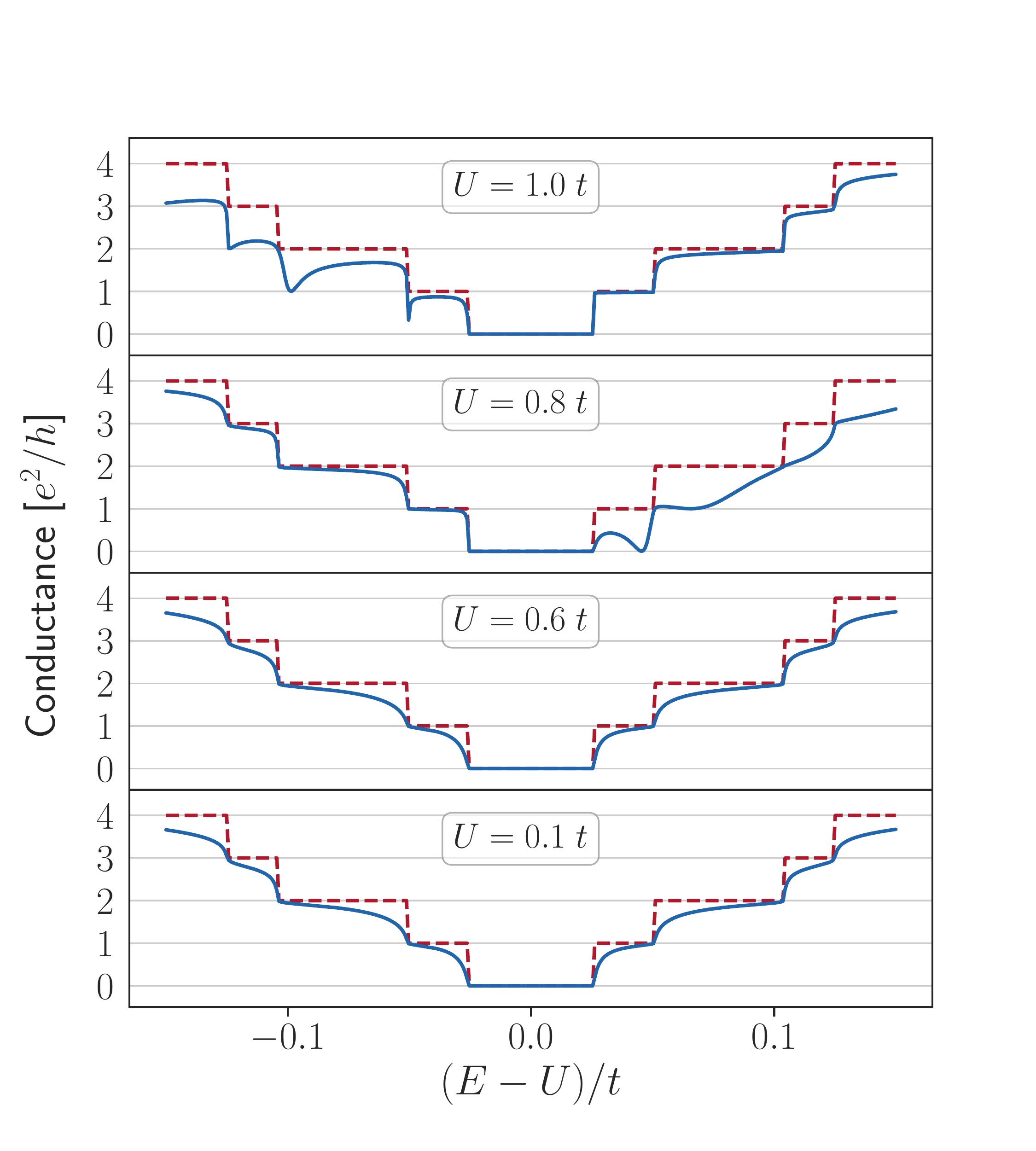}
    \vspace*{-0.55cm}
    \caption{Conductance expressed in terms of $e^2/h$ as a function of $(E-U)/t$ for a aGNR of width $ W \sim 12 \, \mathrm{nm}$ and length $L\sim 15 \, \mathrm{nm}$ for several values of $U$ expressed in terms of $t$. Dashed lines represent the conductance of pristine ribbons and continuous lines show the conductance for ribbons with a vacancy at $(x,y)\approx (L/2,W-1.5a)$.}
    \label{fig:accon2}
\end{figure}

\subsection{TBH regime} \label{sec:TBH}

\begin{figure}[ht]
    \vspace*{-1cm}
    \centering
    \includegraphics[width=0.8\columnwidth]{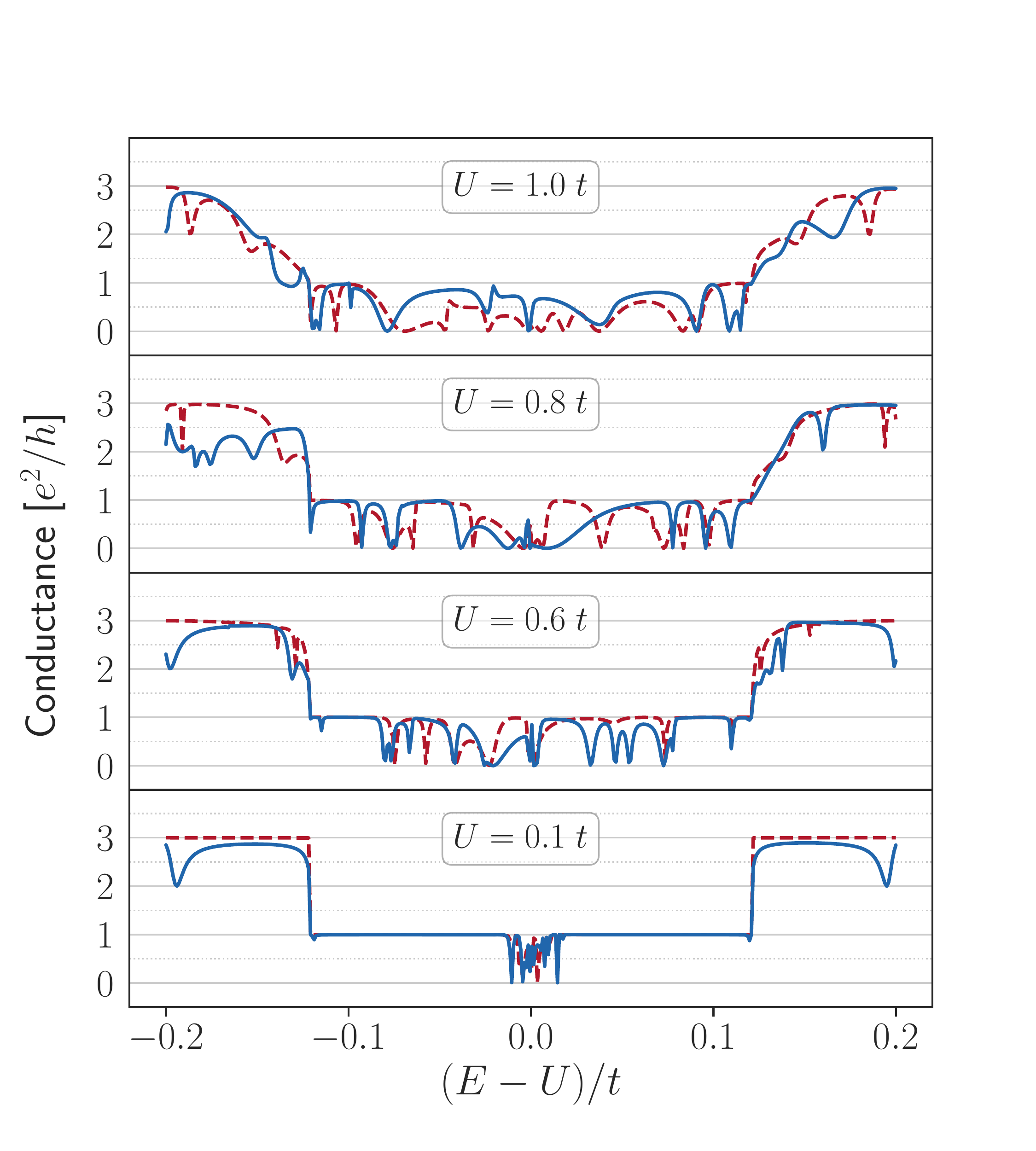}
    \vspace*{-0.55cm}
    \caption{Conductance expressed in terms of $e^2/h$ as a function of $(E-U)/t$ for a zGNR of width $ W \sim 12\,\mathrm{nm}$ and length $L\sim 15\,\mathrm{nm}$ for several values of $U$ expressed in terms of $t$. Dashed lines represent the conductance of pristine ribbons and continuous lines show the conductance for GNRs with a vacancy located at $(x,y)\approx (L/2,W-1.5a)$.}
    \label{fig:zzcon2}
     %\vspace*{-0.85cm}
\end{figure}

In order to understand the impact of $e-e$ interactions on the electronic transport in \textit{trivial} GNRs, in Figs.~\ref{fig:accon2} and~\ref{fig:zzcon2} we show the conductance, in terms of $e^2/h$, as a function of energy for aGNRs and zGNRs in the TBH regime for several values of $U$.

\begin{figure}[ht]
    %\vspace*{-0.75cm}
    \centering
    \includegraphics[width=0.8\columnwidth]{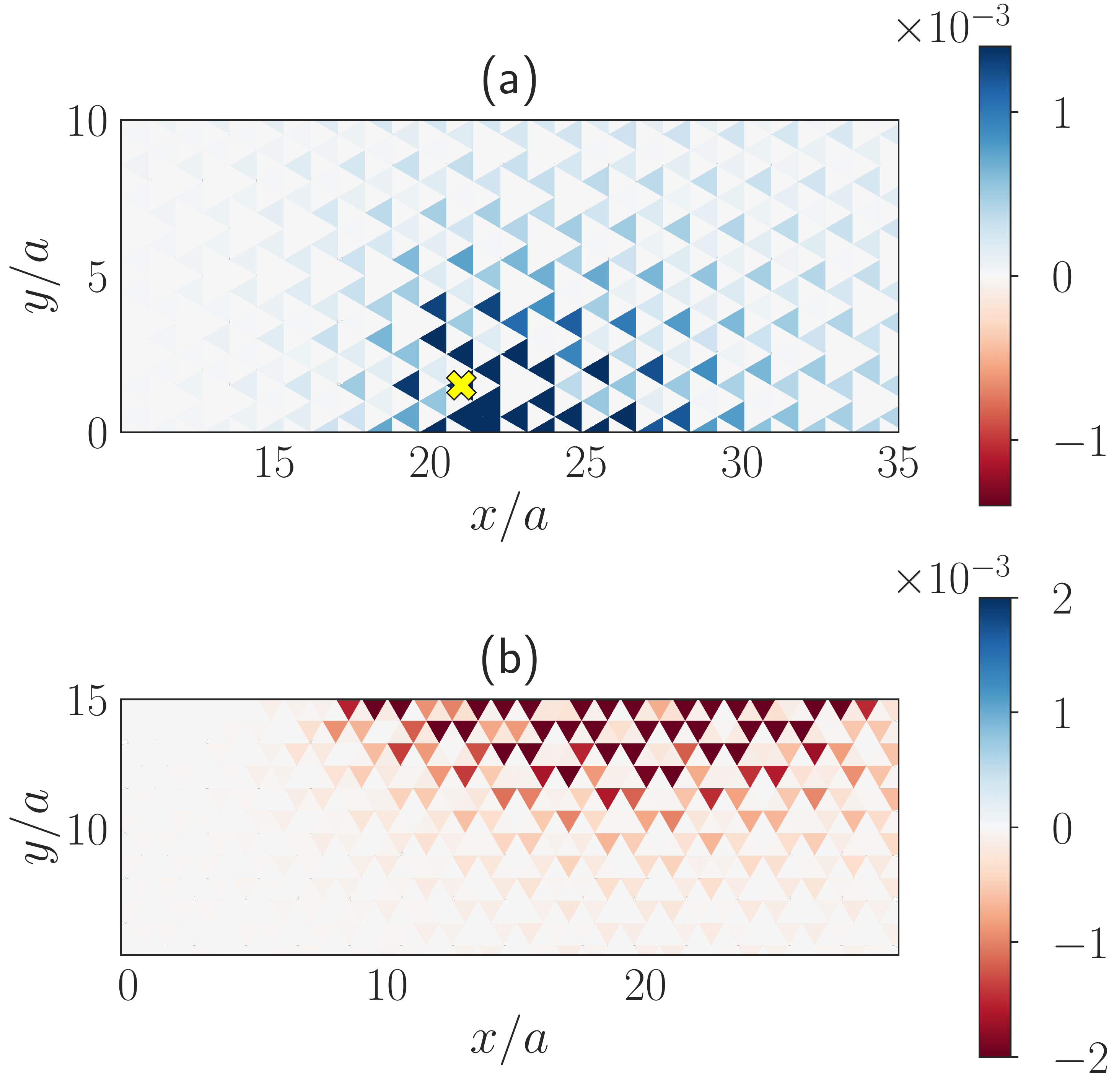}
    \caption{Dimensionless $\hat{x}$  ($m_x$) component in TBH regime obtained from Eq.~(\ref{eq:eig}) in (a)~a defective aGNR with $U/t=0.8$ and a vacancy at $(x,y)=a(21,1.5)$ which is indicated with a yellow $\bm{\times}$ at the resonant energy of the dip shown in Fig.~\ref{fig:accon2}, (b)~a pristine zGNR with $U/t=0.6$ at an energy where $\tau = 0$, e.g. $E \sim 0.68t$. $m_z$ is negligible and $m_y$ is similar to panel~(a) but with opposite sign.}
    \label{fig:l0comp}
\end{figure}

The results shown in Fig.~\ref{fig:accon2} correspond to insulating aGNRs of width $W\sim 11\,\mathrm{nm}$ and length $L\sim 15\,\mathrm{nm}$. In this regime, the transmission coefficient near Fermi energy is $\tau=0$ as expected for an insulating aGNR. Dashed curves, corresponding to pristine aGNRs, show that Coulomb interactions do not affect the transport properties of pristine aGNRs, which is consistent with the results for the QSH regime. In the case of defective aGNRs, the effect of the vacancy is the same as expected for a typical TB model for small values of $U$ ($U/t \sim 0.1,0.6$) namely, a small deviation from perfect conductance due to the formation of a quasi-bound state localized at the vacancy arising from an enhancement of the LDOS around it.  For larger values of $U$, the conductance shows resonant dips at the threshold of the second and subsequent modes, similar to the ones arising in zGNRs with vacancies near the edge when $e-e$ interactions are not considered~\cite{Xiong2010}. In Fig.~\ref{fig:l0comp}(a) we plot the $x$-component of the magnetic moment for a defective aGNR with $U/t=0.8$ at the dip resonant energy. From this profile we can thus associate these resonant dips to the formation of a quasi-bound state localized around the vacancy whose resonant energy is proportional to $U$. The main difference between TBH and TB regime is that in the case of TBH these localized states induce a nontrivial in-plane magnetization through an enhancement of the LDOS. Whereas, when compared to QSH regime, the main difference is that the localized bound state does not trigger spin-flip events due to the absence of helical edge states.

The results shown in Fig.~\ref{fig:zzcon2} correspond to zGNRs of width $W\sim 12 \; \mathrm{nm}$ and length $L\sim 15\,\mathrm{nm}$. In this regime, the nondispersive edge modes are the responsible of transport at low energies. The mechanism behind electric conduction relies in the exponential decay of these localized modes into the bulk, which allows the transmission of a mode which current density peaks at the center of the GNR~\cite{Zrbo2007}. Hence, by studying transport in TBH regime we can ascertain the effects of $e-e$ interactions in these nondispersive modes and gain insight of transport in QSH regime. The most remarkable result of Fig.~\ref{fig:zzcon2} is that $e-e$ interactions spoil conductance quantization in both pristine and defective samples just as in QSH regime. Furthermore, in this case the breakdown of quantization is even greater, leading to complete suppression of transmission at some energies. Both in pristine and defective samples, the transmission coefficient oscillates between $\tau=0$ and $\tau=1$ in an energy interval (which we shall call \textit{resonant interval} from now on), whose width is proportional to $U$, even covering our interval of study completely for $U/t=1.0$. This can also be observed in Fig.~\ref{fig:zzcon}, where it is more subtle. The effect of vacancies in electronic transport is not important: the presence of a vacancy near the edge just changes slightly the pattern of the aforementioned oscillations and induces a resonant dip for $E=U\pm 0.2t$ for low values of $U$, $0.1 \leq U/t \leq 0.6$. In order to understand the mechanism behind these oscillations, we plot the spatial profile of the $x$-component of the magnetic moment in Fig.~\ref{fig:l0comp}(b). For every pristine and defective GNR, at energies lying in the \textit{resonant interval}, the spatial profile for the magnetic moment (or for the LDOS) is similar: it displays a peak in the LDOS, which has a definite nontrivial in-plane magnetic moment and it is located at the edge, while the rest of the GNRs present low to zero LDOS and thus no magnetization. As in the QSH regime, spontaneous magnetization leads to the breakdown of conductance quantization, but in TBH regime this magnetization is more destructive because dispersive helical states are not present. In the TBH regime spin-flip events do not arise, but spontaneous localization of states lead to a partial disappearance of edge states and thus hinders the exponential decay into the bulk [see left panels of Fig.~\ref{fig:l0comp}(b)], spoiling conductance quantization.

\subsection{Thermoelectric efficiency}\label{sec:termo}

After discussing the transmission coefficients for aGNRs and zGNRs in both regimes and for several values of $U$, we apply the formalism presented in Sec.~\ref{sec:Model} in order to obtain the figure of merit $ZT$ for each case. To that end, we computed $ZT$ with  Eq.~(\ref{eq:zt}) for several values of $\mu$ in the range $[U/t-0.2,U/t+0.2]$ so as to determine the maximum values $ZT_{\text{max}}$ and $\mu_{\text{max}}$. These results are presented in Table~\ref{tab:tabla} for each case studied in the previous section. Table~\ref{tab:tabla} shows that the most thermoelectrically efficient systems are the zGNRs in TBH regime, displaying values up to  $ZT_{\text{max}}=2.01$ for the defective ribbon with $U/t=0.6$. The second most thermoelectrically efficient systems are the aGNRs in TBH regime, displaying values of $ZT_{\text{max}} \sim 10^{-2}$. 

\begin{figure}[t!]
\hspace*{-0.9cm}
    %\vspace*{-0.5cm}
    \centering
    \includegraphics[width=0.8\columnwidth]{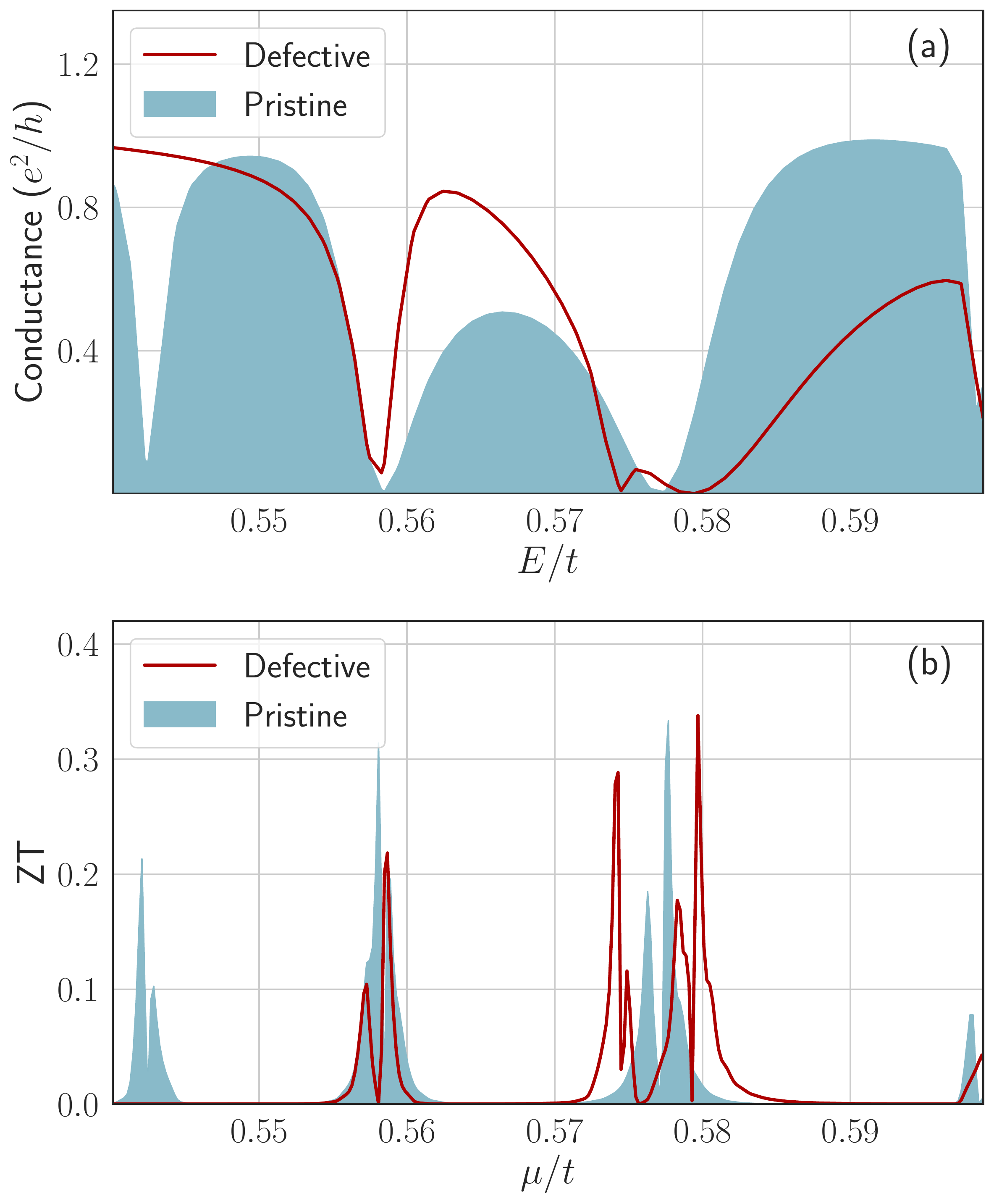}
    \vspace*{-0.3cm}
    \caption{(a) Conductance in terms of $e^2/h$ as a function of energy for a pristine and a defective zGNRs in the TBH regime with $U/t=0.6$. (b)~$ZT$ as a function of the chemical potential for a pristine and defective zGNR in TBH regime with $U/t=0.6$.}
    \vspace*{-0.3cm}
    \label{fig:ZT}
\end{figure}

Meanwhile, in QSH regime, both aGNRs and zGNRs show values of $ZT_{\text{max}} \sim 10^{-3}.$ We can thus study the interplay between vacancies, $\lambda$ and $U$ in thermoelectric efficiency. $ZT_{\text{max}}$ in the QSH regime is lower than $ZT_{\text{max}}$ in the TBH regime. Topological protection of the helical edge modes seem to prevent the marked fluctuations in the transmission coefficient happening in zGNRs in TBH, where the effect of $U$ is most destructive. Surprisingly, the presence of vacancies does not seem to have a huge impact in $ZT_{\text{max}}$, albeit in general, defective samples display slightly greater values of $ZT_{\text{max}}$. The value of $ZT_{\text{max}}$ is rather independent of $U$ but, as shown in the previous section, $e-e$ interactions are crucial for the mechanisms behind the appearance of resonant dips, specially in zGNRs.

% The value of $U$ does not seem to have a huge impact in $ZT_{\text{max}}$ neither in QSH or TBH regime. Nonetheless, as presented in the previous section, in zGNRs the energy interval where interference and hence resonant dips which enhance thermoelectric efficiency can arise is proportional to the value of $U$, relating the general increase of $ZT$ to $U$ in this case. Furthermore, in the case of aGNRs, even though $ZT$ seems virtually independent of $U$, $e-e$ interactions are the responsible of the mechanism behind backscattering. The value of $U$ is, in general, not important for thermoelectric efficiency but the $e-e$ interactions are crucial.

A description of how interference enhances $ZT$ is shown in Fig.~(\ref{fig:ZT}), where asymmetric lineshapes of the transmission coefficient around a certain resonance energy $E_R$ give rise to peaks in $ZT$ at $\mu=E_R$. We can approach this phenomenon in two ways. First, from Sec.~\ref{sec:Model} we know that $ZT \propto S \propto K_1$. In $K_1$, the product $\partial_{E}f\times(E-\mu)$ is an odd function with respect to $E-\mu$ and therefore, an asymmetric lineshape of $\tau$ around $\mu$ will boost $K_1$ and thus $ZT$~\cite{SaizBretn2015}. Second, we shall remind that resonant dips in transmission arise from interference phenomena. This interference takes place between the two possible paths an electron can take while traveling through the scattering region, namely, one in which it takes part in some scattering process and one in which it travels without being scattered. In our work, these scattering processes arise due to nontrivial magnetic moments around vacancies induced by peaks in the LDOS. We argue that the thermoelectric efficiency boosts due to interference works in this way: at the resonant energy $E_R$, the transmission is heavily hindered due to these aforementioned scattering processes. 
A small increase in the electron energy amounts to 
%uplift the possibilities of it bypassing these processes and thus
an enhancement of its transmission. 
Hence, a heat flux through the sample with $\mu=E_R$ would lead to an sizable variation in the electronic transmission, resulting in an increase of the electric current and enhancing the thermoelectric effect. This would explain why asymmetric lineshapes, related with more destructive scattering processes, lead to greater $ZT$ values rather than symmetric lineshapes, which are related to elastic scattering phenomena. Thus,  electron-electron interactions can give rise to interference phenomena which can notably increase the thermolectric response of a nanostructure.

\begin{table*}[ht]
\vspace*{-0.6cm}
\centering
\begin{tabular}{ccccccccc}
\toprule
\toprule
       & \multicolumn{4}{c}{Pristine aGNR}                         & \multicolumn{4}{c}{Pristine zGNR}                                                 \\ 
       & \multicolumn{2}{c}{QSH}     & \multicolumn{2}{c}{TBH}     & \multicolumn{2}{c}{QSH}     & \multicolumn{2}{c}{TBH}                             \\ 
$U/t$  & $ZT_\mathrm{max}$             & $\mu/t$   & $ZT_\mathrm{max}$  & $\mu/t$   & $ZT_\mathrm{max}$ & $\mu/t$   & $ZT_\mathrm{max}$       & $\mu/t$   \\ \cmidrule(r){1-1} \cmidrule(r){2-3} \cmidrule(r){4-5} \cmidrule(r){6-7} \cmidrule(r){8-9}
$\;\;\;0.10\;\;\;$ & $\;\;\;9.52\times10^{-3}\;\;\;$ & $\;\;\;0.17\;\;\;$ & $\;\;\;2.71\times10^{-2}\;\;\;$ & $\;\;\;0.15\;\;\;$ & $\;\;\;7.67\times10^{-3}\;\;\;$ & $\;\;\;0.09\;\;\;$ & $\;\;\;3.67\times10^{-1}\;\;\;$                         & $\;\;\;0.10\;\;\;$ \\
$0.60$ & $9.52\times10^{-3}$ & $0.67$ & $2.71\times10^{-2}$ & $0.65$ & $3.94\times10^{-3}$ & $0.61$ & $3.33\times10^{-1}$                         & $0.59$ \\
$0.80$ & $9.52\times10^{-3}$ & $0.87$ & $2.71\times10^{-2}$ & $0.85$ & $7.27\times10^{-3}$ & $0.93$ & $4.77\times10^{-1}$                         & $0.81$ \\
$1.00$ & $9.52\times10^{-3}$ & $0.93$ & $2.71\times10^{-2}$ & $1.05$ & $2.86\times10^{-4}$ & $1.12$ & $4.30\times10^{-1}$                         & $0.97$ \\ 
\bottomrule
\toprule
       & \multicolumn{4}{c}{Defective aGNR}                     & \multicolumn{4}{c}{Defective zGNR}                                             \\ 
       & \multicolumn{2}{c}{QSH}     & \multicolumn{2}{c}{TBH}     & \multicolumn{2}{c}{QSH}     & \multicolumn{2}{c}{TBH}                             \\ 
$U/t$  & $ZT_\mathrm{max}$             & $\mu/t$   & $ZT_\mathrm{max}$  & $\mu/t$   & $ZT_\mathrm{max}$ & $\mu/t$   & $ZT_\mathrm{max}$       & $\mu/t$   \\ \cmidrule(r){1-1} \cmidrule(r){2-3} \cmidrule(r){4-5} \cmidrule(r){6-7} \cmidrule(r){8-9}
$0.10$ & $7.72\times10^{-3}$ & $0.13$ & $1.12\times10^{-2}$ & $0.10$ & $3.35\times10^{-3}$ & $0.23$ & $2.53\times10^{-1}$                         & $0.11$ \\
$0.60$ & $8.05\times10^{-3}$ & $0.63$ & $1.13\times10^{-2}$ & $0.60$ & $3.48\times10^{-3}$ & $0.73$ & $2.01$                                     & $0.60$ \\
$0.80$ & $8.30\times10^{-3}$ & $0.77$ & $3.60\times10^{-1}$ & $0.80$ & $1.97\times10^{-3}$ & $0.86$ & $7.12\times10^{-1}$ & $0.80$ \\
$1.00$ & $8.51\times10^{-3}$ & $0.93$ & $8.49\times10^{-3}$ & $0.95$ & $4.39\times10^{-3}$ & $1.11$ & $5.67\times10^{-1}$                        & $0.99$ \\ \bottomrule \bottomrule
\end{tabular}
\caption{Maximum value of $ZT$ and the value of $\mu$ at which it arises for the aGNRs and zGNRs studied in this work.}
\label{tab:tabla}
\vspace*{-0.4cm}
\end{table*}

\section{Conclusions and Experimental proposal}\label{sec:concl}

In summary, in this work we studied the transport properties of aGNRs and zGNRs in QSH and TBH regimes in order to determine the effect of $e-e$ interactions in their thermoelectric efficiency. We found that $e-e$ interactions are crucial for the appearance of the interference phenomena that give rise to the enhanced of the thermoelectric efficiency in the GNRs considered in this work. Even so, $ZT_{\text{max}}$ is, in general, unaltered by the value of $U$ and mainly depends on the regime and edge termination of the GNR. For pristine aGNRs in TBH regime and defective aGNRs in QSH regime, $ZT_{\text{max}}$ is virtually independent of $U$. Meanwhile, in defective aGNRs in TBH regime, it is most important since resonant dips only appear for large values of $U$. For pristine and defective zGNRs, even though the value of $ZT_{\text{max}}$ is not influenced by $U$, large values of $U$ lead to larger intervals of energy where transmission coefficient oscillates abruptly and thus enhances $ZT$ in a wider range of chemical potentials $\mu$. 
% We have thus carried out a detailed study of the interplay between $e-e$ interactions, topological dispersive edge modes and geometric non-dispersive edge modes in thermoelectric efficiency in what we can understand as toy models of 2DTIs and 2D trivial insulators. 

The experimental implementation of this work is feasible. We propose the fabrication of encapsulated graphene nanoconstrictions in a Hall bar configuration with well defined edges (aGNR and zGNR) through the cryo-etch method introduced in Ref.~\cite{Cleric2019}. In order to obtain the QSH regime, in-plane magnetic field is suggested according to Ref.~\cite{Young2013}. Meanwhile, modulation of $e-e$ interactions is proposed following Ref.~\cite{Kim2020}. We propose nonlocal measurements in order to test the effect of $e-e$ interactions in the helical edge states and the possible application of an electric field as in Ref. ~\cite{SaizBretn2015} to enhance figure of merit $ZT$. Low temperatures are intended ($T\sim 4$ K) but higher temperature experiments are recommended in future works in order to understand the lattice contribution to $ZT$, as studied in Ref.~\cite{SaizBretn2019}. 

\appendix

\section{Mean-field approximation} \label{sec:MFA}

Due to the many-body nature of the Hamiltonian (\ref{eq:ham1}), its exact solution is rather unmanageable. In order to obtain an effective one-particle Hamiltonian we apply an unrestricted Hartree-Fock mean-field approximation to deal with the Hubbard term~\cite{Tosatti2006,Vergs1992}. Within this approximation, we consider fluctuations in the creation-annihilation pairs as
\begin{align} \label{eq:Fock1}
\begin{split}
    n_{i,\alpha}n_{j,\beta} &= c_{i,\alpha}^{\dagger}c_{i,\alpha}^{}c_{j,\beta}^{\dagger}c_{j,\beta}^{} \approx c_{i,\alpha}^{\dagger}c_{i,\alpha}^{} \langle c_{j,\beta}^{\dagger}c_{j,\beta}^{} \rangle\\
    &  + \langle c_{i,\alpha}^{\dagger}c_{i,\alpha}^{} \rangle  c_{j,\beta}^{\dagger}c_{j,\beta}^{} - \langle c_{i,\alpha}^{\dagger}c_{i,\alpha}^{}c_{j,\beta}^{\dagger}c_{j,\beta}^{}  \rangle\ .
    \end{split}
\end{align}
We choose this approximation in the interest of studying in-plane spin-flip events that do not appear when considering a Hartree approximation, which decouples the number operator $n_{i,\alpha}$ rather than creation-annihilation pairs. Due to Wick's theorem, considering all pairs of operators, from Eq.~(\ref{eq:Fock1}) we determine that
\begin{align} \label{eq:Fock2}
\begin{split}
    n_{i,\uparrow}n_{i,\downarrow} &\approx \langle c_{i,\uparrow}^{\dagger}c_{i,\uparrow}^{} \rangle c_{i,\downarrow}^{\dagger}c_{i,\downarrow}^{} + c_{i,\uparrow}^{\dagger}c_{i,\uparrow}^{}\langle c_{i,\downarrow}^{\dagger}c_{i,\downarrow}^{} \rangle \\ &- \langle c_{i,\uparrow}^{\dagger}c_{i,\uparrow}^{} \rangle \langle c_{i,\downarrow}^{\dagger}c_{i,\downarrow}^{} \rangle-\langle c_{i,\uparrow}^{\dagger}c_{i,\downarrow}^{} \rangle c_{i,\downarrow}^{\dagger}c_{i,\uparrow}^{}\\
    &-c_{i,\uparrow}^{\dagger}c_{i,\downarrow}^{} \langle c_{i,\downarrow}^{\dagger}c_{i,\uparrow}^{} \rangle +\langle c_{i,\uparrow}^{\dagger}c_{i,\uparrow}^{} \rangle\langle c_{i,\downarrow}^{\dagger}c_{i,\uparrow}^{} \rangle\ .
\end{split}
\end{align}
After rearranging the terms of Eq. (\ref{eq:Fock2}), and defining $S_{i}^{+} = c_{i,\uparrow}^{\dagger}c_{i,\downarrow}^{}$ and $S_{i}^{-} = c_{i,\downarrow}^{\dagger}c_{i,\uparrow}^{}$, we can express the Hamiltonian~(\ref{eq:ham1}) approximately as
\begin{align} \label{eq:ham2}
    \begin{split}
            &\mathcal{H}_{\mathrm{MF}} \approx \; \mathcal{H}_0+U\sum_i\left( \langle n_{i\downarrow}\rangle n_{i\uparrow}+\langle n_{i\uparrow}\rangle n_{i\downarrow} - \langle n_{i\uparrow}\rangle \langle n_{i\downarrow}\rangle \right)\\
    &-U\sum_i\left(\langle S_i^{+}\rangle c_{i\downarrow}^{\dagger}c_{i\uparrow}^{}+\langle S_i^{-}\rangle c_{i\uparrow}^{\dagger}c_{i\downarrow}^{}- \langle S_i^{+}\rangle\langle S_i^{-}\rangle \right)\ ,,
    \end{split}
\end{align}
where $\mathcal{H}_0$ stands for the noninteracting part of the Hamiltonian (\ref{eq:ham1}). \

The $\langle S_{i}^{\pm} \rangle$ and $\langle n_{i,\alpha} \rangle$ terms need to be obtained self-consistently. In order to do so, we will work in the basis where the lattice sites are ordered like 
$$
\{1^A_{\uparrow},1^A_{\downarrow},1^B_{\uparrow}, 1^B_{\downarrow}, 2^A_{\uparrow},\dotsm,(N/2)^B_{\downarrow}\}\ ,
$$ 
where $N$ is the number of sites of the lattice and A and B are the two sublattices of the honeycomb lattice. We employ the following algorithm~\cite{Feldner2010}
\begin{enumerate}

    \item Generate a random set of $4$ vectors of length $N$,
    \begin{align} \label{eq:set}
        \mathcal{V}_0 = \{\langle S_i^{+}\rangle,\langle S_i^{-}\rangle,\langle n_{i,\uparrow}\rangle,\langle n_{i,\downarrow}\rangle\}_{i=1}^{N}\ .
    \end{align}
    A random first choice may slow down convergence but helps to avoid magnetic frustration of the system.
    
    \item From the set (\ref{eq:set}) we build the $2N\times 2N$ matrix representation of the Hamiltonian (\ref{eq:ham2}) and diagonalize it 
    \begin{align}
        \hat{\mathcal{H}}_{\mathrm{MF}}(\mathbf{k})\psi_{\alpha}(\mathbf{k})= \epsilon_{\alpha}(\mathbf{k})\psi_{\alpha}(\mathbf{k})\ ,
    \end{align}
    where $\psi_{\alpha}(\mathbf{k})=(\psi_{\alpha}^{1A},\psi_{\alpha}^{1B},\dotsm,\psi_{\alpha}^{NB})^T$. In this step we obtain $2N$ eigenvalues $\{\epsilon_v \}_{v=1}^{2N}$ and $2N$ $2$-component eigenvectors $\phi_v$ of the form
    \begin{align}
        \phi_{v} = \begin{bmatrix}
        \phi_{v,\uparrow} \\
        \phi_{v,\downarrow} 
        \end{bmatrix},
    \end{align}
    where $v$ is the band index and $\phi_{v}$ is defined to be $\psi_{\alpha}(\mathbf{k})=\{\phi_{v,\alpha}\}_{v=1}^N$.
    
    \item From the eigenvectors and eigenvalues of step 2, we can obtain a new set $\mathcal{V}_1$ analogous to the one of Eq.~(\ref{eq:set}) by computing
    \begin{align} \label{eq:eig}
        &\langle \mathbf{m}_i\rangle^{(1)} = \sum_{v=1}^{N_e}[\phi_{v,i}^{(0)}]^{\dagger}\; \bm{\sigma} \; [\phi_{v,i}^{(0)}]^{}\ , \\
        &\langle n_{i,\uparrow} \rangle^{(1)} = \sum_{v=1}^{N_e}[\phi_{v,i}^{(0)}]^{\dagger} \begin{pmatrix}
      1 & 0 \\
      0 & 0 
      \end{pmatrix}[\phi_{v,i}^{(0)}]^{}\ ,\\
      &\langle n_{i,\downarrow} \rangle^{(1)} = \sum_{v=1}^{N_e}[\phi_{v,i}^{(0)}]^{\dagger} \begin{pmatrix}
      0 & 0 \\
      0 & 1 
    \end{pmatrix}[\phi_{v,i}^{(0)}]^{}\ ,
    \end{align}
    where $\bm{\sigma}$ is a three-component vector formed by the Pauli matrices and $\langle \mathbf{m}_i\rangle$ is the magnetic moment of the $i$-th site. In the above expressions we considered a fixed number of electrons $N_e$ in the system, fixing right away the chemical potential $\mu$ like $\epsilon_{N_e}=\mu$. In this work we consider half-filling, setting $N_e=N$. From Eq.~(\ref{eq:eig}) we determine $\langle S_i^{\pm}\rangle$ as
    \begin{align}
        \langle S_i^{\pm} \rangle = \frac{\langle m_x \rangle \pm i\langle m_y \rangle}{2}\ .
    \end{align}
    
    \item  With the new set $\mathcal{V}_1$, repeat steps 2 and 3 iteratively until convergence. Convergence for the $k$-th iteration is defined as
    \begin{align}
        \{\mathcal{V}_k^{(n)}(i)\}_{n=1}^4 - \{\mathcal{V}_{k-1}^{(n)}(i)\}_{n=1}^4 \leq \mathcal{T} \ , \forall i\ ,
    \end{align}
    where $\mathcal{T}$ is the tolerance, $i$ the lattice site and $n$ the component of $\mathcal{V}_k$. In this work we used a variable weight method, using information from iteration $k-1$ and $k-2$ for iteration $k$ as follows
    \begin{align}
    \begin{split}
       \{\mathcal{V}_k^{(n)}(i)\}_{n=1}^4 &= \left(1-\frac{k}{k^{1.5}+1}\right)\\
       \times \{\mathcal{V}_{k-2}^{(n)}(i)\}_{n=1}^4 &+\frac{k}{k^{1.5}+1}\{\mathcal{V}_{k-1}^{(n)}(i)\}_{n=1}^4\ ,
     \end{split}
    \end{align}
    where we impose that $k/\left(k^{1.5}+1\right) \geq \mathcal{T}$. This weight method helps to avoid stagnation and hence accelerates convergence.
\end{enumerate}

\begin{figure}[ht]
    \centering
         \includegraphics[width=0.8\columnwidth]{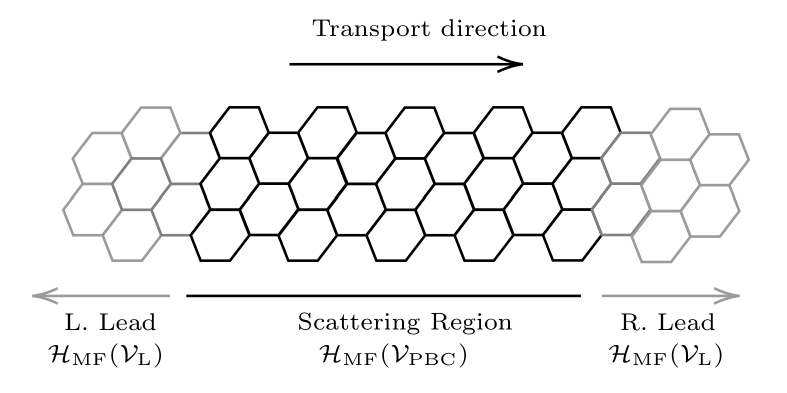}

    \caption{Schematic diagram of the scattering region and the leads used in the transport calculations.}
    \label{fig:transport}
\end{figure}

Transport properties at low temperature were calculated within the Landauer - Büttiker formalism~\cite{Datta1995,Davies1997} as implemented in the {\tt Kwant} toolkit~\cite{Groth2014}. In this formalism, conductance at energy $E$ is defined as 
\begin{align}\label{eq:landauer}
    G(E)=\frac{e^2}{h}\tau (E),
\end{align}
where $e$ is the electron charge, $h$ is the Planck's constant and $\tau (E)$ is the transmission coefficient at that energy. Electronic transport is thus understood to be a scattering phenomenon with a statistical interpretation, where two pristine leads, connected to two reservoirs, are connected to a scattering region as shown in Fig.~\ref{fig:transport}. Conductance at energy $E$ is directly proportional to the modes which can travel along the scattering region.

In order to study the scattering phenomenon, we applied the selfconsistent algorithm to a GNR with periodic boundary conditions along the transport direction, and obtained the set $\mathcal{V}_{\mathrm{PBC}}$. Afterwards, we defined a scattering region of the same size with Hamiltonian $\mathcal{H}_{\mathrm{MF}}({\mathcal{V}_{\mathrm{PBC}}})$ and attached the two leads to the system with Hamiltonian $\mathcal{H}_{\mathrm{MF}}({\mathcal{V}_{\mathrm{L}}})$ where $\mathcal{V}_{\mathrm{L}}= \{ 0,0,0.5,0.5 \}$ for every lattice site $i$. Once the device is prepared, we compute the conductance using the scattering matrix formalism~\cite{Datta1995}.

\acknowledgments

This work was supported via Grants PID2019-106820RB-C21 and PGC2018-097018-B-I00 (MCIN/ AEI/FEDER, EU). The authors are grateful to Enrique Diez and Mario Amado for helpful discussions.

\end{document}